\documentclass[twocolumn,prb,aps,superscriptaddress]{revtex4}
\usepackage{color}
\usepackage{graphicx}% Include figure files
\usepackage{tabularx}
\usepackage{natbib}
\usepackage{graphicx}
\usepackage{bm}
\usepackage{epsf}
\usepackage{epstopdf}
\usepackage{hyperref}
\usepackage{longtable}
\usepackage{url}
\usepackage{amsmath}
\usepackage{amssymb}
\usepackage{amsxtra}
\usepackage{amscd}
\usepackage{amsthm}
\usepackage{amsfonts}
\usepackage{eucal}
\usepackage{latexsym,amsthm}
\usepackage{pstricks}
\usepackage{subfigure}

\newcommand{\nc}{\newcommand}
\nc{\on}{\operatorname}
\nc{\wt}{\widetilde}
\nc{\Wick}{{\mathbb :}}
\nc{\R}{{\mathbb R}}

\newcommand{\beq}{\begin{equation}}
\newcommand{\eeq}{\end{equation}}
\newcommand{\bmul}{\begin{multline}}
\newcommand{\emul}{{\end{multline}}}
\newcommand\beqa{\begin{eqnarray}}
\newcommand\eeqa{\end{eqnarray}}
\newcommand\bea{\begin{array}}
\newcommand\eea{\end{array}}
\newcommand\ba{\begin{array}}
\newcommand\ea{\end{array}}

\newcommand{\neqa}{\nonumber\end{eqnarray}}

\nc{\CH}{{\mathcal H}}
\nc{\Db}{{\bar D}}
\nc\comment[1]{}

\nc{\CM}{{\mathcal M}}
\nc{\CN}{{\mathcal N}}

\renewcommand{\v}{{\rm v}}

\newcommand{\re}{\relax{\rm I\kern-.18em R}}

\nc{\meV}{{\mathrm{\,meV}}}
\nc{\cG}{{\mathcal G}}

\renewcommand{\)}{\right)}

\comment{ \)}
\renewcommand{\bar}{\overline}

\nc{\al}{{\alpha}}

\begin{document}
\title{Entanglement scaling and spatial correlations of the transverse field Ising model with perturbations.}
\author{Richard Cole}
%\email{R.Cole@lboro.ac.uk}
\affiliation{Department of Physics,  Loughborough University, Loughborough LE11 3TU, UK}
\author{Frank Pollmann}
%\email{frankp@mpipks.mpg.de}
\affiliation{Max Planck Institute for Physics of Complex Systems, Noethnitzer Strasse, Dresden , Germany}
\affiliation{Department of Physics, Technical University of Munich, 85748 Garching, Germany}
\author{Joseph J. Betouras}
\email{J.Betouras@lboro.ac.uk}
\affiliation{Department of Physics,  Loughborough University, Loughborough LE11 3TU, UK}
%\pacs{75.70.Ak , 73.22.Pr}
\begin{abstract}
We study numerically the entanglement entropy and spatial correlations  of the one dimensional transverse field Ising model with three different perturbations. First, we focus  on the out of equilibrium, steady state with an energy current passing through the system. By employing 
a variety of matrix-product state based methods,
%and the infinite time evolved block decimation technique, 
we confirm the phase diagram and compute the entanglement entropy. 
Second, we  consider a small perturbation that takes the system away from integrability and calculate
%using the infinite Density Matrix Renormalization Group (iDMRG) 
the correlations, the central charge and the entanglement entropy. 
Third, we  consider periodically weakened bonds, exploring the phase diagram and entanglement properties first in the situation when the weak and strong bonds alternate (period two-bonds) and then the general situation of a period of $n$ bonds. In the latter case we find a critical weak bond that scales with the transverse field as $J'_c/J = (h/J)^n$, where $J$ is the strength of the strong bond, $J'$ of the weak bond and $h$ the transverse field. We explicitly show that the energy current is not a conserved quantity in this case.

\end{abstract}
\maketitle

\section{Introduction}
The transverse field Ising model (TFIM) possesses a central role in both quantum statistical and condensed matter physics.\cite{Dutta} It is used as the benchmark model where new concepts, ideas and theoretical techniques have been derived from or tested. With the surge of activity on quantum information, this model also plays a very important role in simulating interactions among qubits  \cite{Frontiers, Qureshi} and serves as a quantum paradigm which can be explored by novel means.
While the pure TFIM can be solved exactly \cite{Pfeuty, Lieb}, relevant physical perturbations (such as longitudinal fields or spin-exchange)  prohibit an exact solution. In this case advanced numerical tools have to be employed to understand the physics of the models. 
%At the same time, new computational methods have been developed in the past few years and have led to the computations of properties, difficult to be computed before. 
In this work, we study the TFIM with ferromagnetic nearest-neighbour interactions with three different perturbations.  
We use a combination of matrix-product state (MPS) based numerical techniques starting with the time evolving block decimation (TEBD) that allows an efficient time-evolution of matrix-product states in real or imaginary time.\cite{Vidal1} 
This method is similar in spirit to the density matrix renormalisation group (DMRG) method that we also employ \cite{White93,Schollwoeck,McCulloch, Kjall} and has been shown to work  for the entanglement spectrum near criticality in finite quantum spin chains \cite{DeChiara}. Ground state entanglement for example of the XY and Heisenberg models shows the emergence of universal scaling behavior at quantum phase transitions. Entanglement is thus controlled by conformal symmetry. Away from the critical point, entanglement gets saturated by a mass scale \cite{Latorre}. Here, we investigate the entanglement, as well as correlations, of the TFIM with perturbations.
The simulations are performed directly in the thermodynamic limit assuming translational invariance with respect to a unit cell.

%For efficiency in parts of this work we also employ directly the infinite-size DMRG (iDMRG) \cite{McCulloch, Kjall}. The key point is that in iDMRG, the Suzuki-Trotter decomposition is not required, leading to a faster implementation.

We start in Sec. II by revisiting the exactly solvable TFIM in a non-equilibrium steady state, with an energy current passing through the system. We confirm the phase diagram  obtained in Refs.~[\onlinecite{Antal, Eisler1}]  
In addition, we provide new insights for the chiral order parameter, the spin-spin correlations and the entanglement properties. In the case of central charge, we found the change from c=1/2 to c=1 as analytically calculated in Ref.~[\onlinecite{Kadar}].
In Sec. III, we add a perturbation that breaks the integrability of the model, seeking to characterise the system using the bipartite entanglement entropy and central charge. 
%In fact, we calculate the phase diagram of the system through the central charge.
%
% both with and without energy current.
Subsequently, in Sec. IV, we weaken the strength of a bond periodically in space, studying the properties of the system first in equilibrium. The critical value of the weak bond scales with the transverse field as $J'_c/J = (h/J)^n$, when $n$ is the spatial period of the weakened bond. We show that the energy current is not a conserved quantity in this case, so there is no out of equilibrium steady state in the sense as in the homogeneous system. We finally summarise and conclude in Sec. V.

%By altering or substituting parts of the chain, a multitude of possible impurities can be imagined. Similar studies have been performed previously on the Heisenberg antiferromagnetic model \cite{Eggert}. Our results gain insight into further problems involving a non-equilibrium steady state, obtaining a numerical solution with iTEBD. Here we consider interactions over nearest neighbours only, neglecting next-nearest-neighbour or longer ranged interactions, i.e. the nature of the chain is governed entirely by one and two site operators.

 %A critical magnetic field is known to display a second-order phase transition within the transverse field Ising model (TFIM) at zero temperature \cite{Pfeuty}. This separates two distinct ordered and disordered regions. Further to this, other work carried out by inducing a non-equilibrium environment introduces a third region beyond a critical current \cite{Antal}, as seen in Fig.\ref{fig:boundary} the influence of impurity effects on these boundaries is of key concern.

%We show that the Ising model with a single sufficiently weakened bond with finite magnetic field (below critical, $h<h_c$) undergoes a phase transition. Here the process of severing the link evolves the system towards a spin singlet state where the magnetic field takes precedence as the dominating energy term.

\section{TFIM with energy current}
\subsection{Diagonalization of Hamiltonian}
The Hamiltonian for the TFIM with energy current is given by
	\begin{eqnarray}
		H&=&H_{Is}+\lambda_1 J^E_{Is}
		\label{Ising_wc}\\
    	H_{Is} &=&-\sum_{i=1}^N J \sigma_i^z \sigma_{i+1}^z + h \sigma_i^x\\
		J^E_{Is}&=&-\sum_{i=1}^N \frac{J h}{2}(\sigma_{i}^z \sigma_{i+1}^y - \sigma_{i}^y \sigma_{i+1}^z)
	\end{eqnarray}

\noindent where $J$ is the strength of the nearest-neighbor interaction, $\vec{\sigma}$ are the Pauli matrices, $h$ is the transverse applied field, $J_E$ the energy current and $\lambda_1$ is a Lagrange multiplier. %The fact that the transverse field term does not commute with the spin-spin interaction term, introduces quantum dynamics in the system \cite{Pfeuty}. 
The energy current $J^E_{Is}$ is derived from the continuity equation for the conserved local energy operator \cite{Zotos} (the derivation is presented in Appendix B).

This Hamiltonian is exactly solvable through a standard process by mapping it to a quadratic fermionic model. 
%that involves a Jordan-Wigner transformation in real space, then a Fourier transformation to momentum space and finally a Bogoliubov transformation \cite{Lieb}. 
After a Bogoliubov transformation and Fourier transform to momentum space, the Hamiltonian in terms of fermionic operators $\gamma_k$ reads:
	\begin{equation}
		H=\sum_k \epsilon_k \left(\gamma^{\dagger}_k \gamma_k - \frac{1}{2}\right)
	\end{equation}
with energy relation \cite{Antal}
	\begin{equation}
		\epsilon_k=\frac{1}{2}\left(\sqrt{J^2+h^2+2Jh \cos{ka}}+L\sin{ka}\right).
	\end{equation}
It is convenient to use the notation $L=Jh\lambda_1$. For small $L\leq J$, the energy spectrum is gapped and no current flows. Then the zero-current phase is extended until $L>J$ where the imposed current density is strong enough to destroy the gap and mix the excited states with the ground state. Then we enter the region of finite energy current flow.
The phase boundaries are calculated by requiring the dispersion relation and its derivative to vanish with respect to the wavenumber $k$ \cite{Antal}. The values of the critical $L_c$ are: $L_c= h$ if $h \geq J$,  or $J$ if $h < J$;  the energy current is non-zero for values of $L \geq L_c$.

\begin{figure}[tp]
	{\includegraphics[width=8cm]{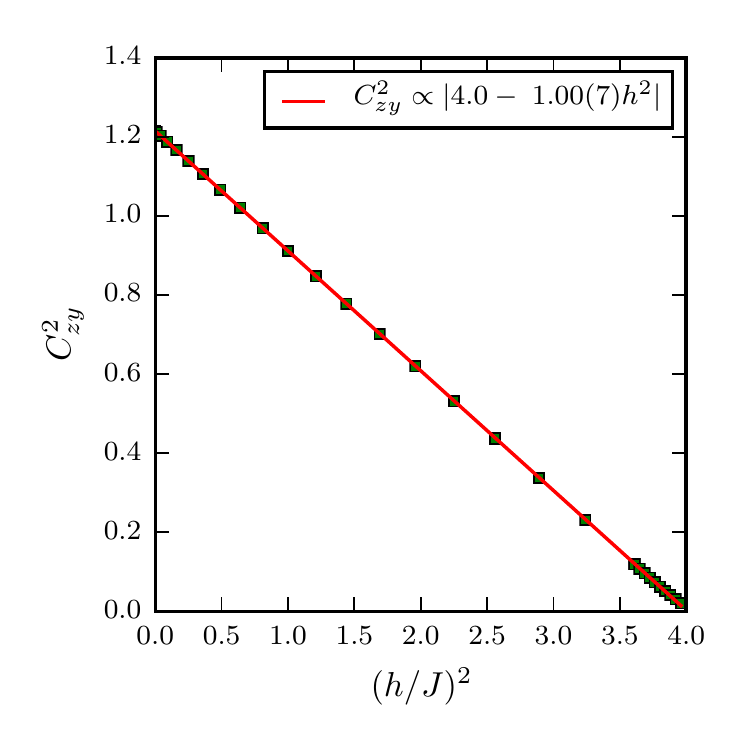}}
	\centering
	\vspace{-0.3 cm}
	\caption{\label{fig:chiral_field} Chiral order parameter $C_{zy}$ as a function of magnetic field in the TFIM with the coefficient of the energy current term $L=2J$ and critical value of the transverse field $h_c=2J$.}
\end{figure}

%\subsubsection{Chiral order}
The transition into the current carrying region can be seen directly through the computation of the chiral order parameter, essentially the average energy current over all bonds:
\begin{equation}
	C_{zy}=\sum_n \langle \sigma^z_n \sigma^y_{n+1} - \sigma^y_n \sigma^z_{n+1}\rangle.
\end{equation}
The expectation value of $C_{zy}$ is: 
\begin{equation}
C_{zy}/N = \frac{1}{\pi L^2} (L^2-h^2)^{1/2} (L^2-J^2)^{1/2}.
\end{equation}
At the phase boundaries, the critical behavior of $C_{zy}$ is: $C_{zy} \propto (L-h)^{1/2}$ if $h  \neq J$ and $C_{zy} \propto (L-J)$ if $h=J$.
This has been also verified by the numerical calculations in the present study. For example $C_{zy}$ as a function of h is shown in Fig. \ref{fig:chiral_field}.

%
%Concerning the dependence of the chiral order on the magnetic field strength, the current is overcome as the model moves into the paramagnet state under a strong magnetic field $h_{c}>L$. Fig.\ref{fig:chiral_field} shows $C_{zy}$ for constant current $L=2J$ as the PM boundary is approached. $C_{zy}$ then as a function of transverse field is given by:
%\begin{equation}
%	\vspace{-5pt}
%	C_{zy}\propto |h^2_{c}-h^2|^{1/2}
%\end{equation}
%extracted from the linear plot in Fig.\ref{fig:chiral_field}.
%
%

\subsection{Spin-spin Correlations}
\begin{figure}[htp]
	{\includegraphics[width=8cm]{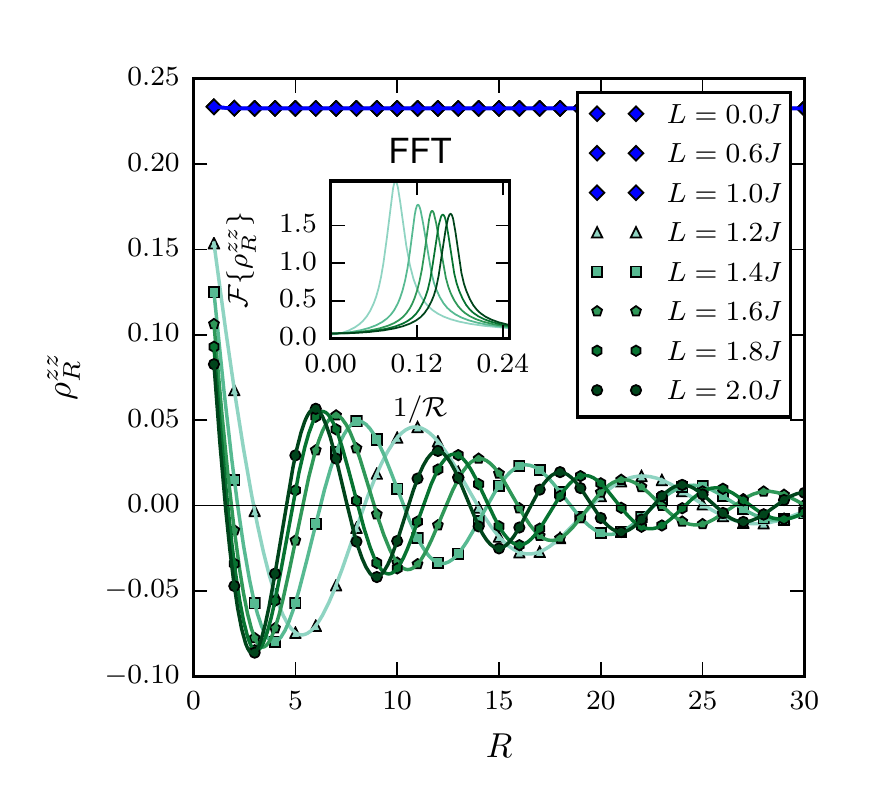}}
	\vspace{-0.3 cm}
	\caption{\label{fig:FMIsing} Spin-spin correlations of the TFIM with energy current $L$ in range $\{0,0.2,...,1.8,2.0\}$ and magnetic field $h=0.5 J$ below the critical value $h_{c}=J$. The oscillatory behavior of the $zz$-correlation function is shown above critical value of the energy current $L_c=J$. In the region $L\leq J$ normal FM correlations are observed. Inset: A fast Fourier transform (FFT) is used to calculate the period as it shifts with $L$.}
\end{figure}

%\subsubsection{Spin-spin correlation function}
The behavior of the spin-spin correlation function $\langle \sigma^z_n \sigma^z_{n+R} \rangle$ reveals distinct properties within the current region ($L=2J$ in Fig. \ref{fig:FMIsing}). We confirm the oscillatory behavior accompanied by a power-law decay (as $1/\sqrt{R}$), where $R$ is the distance (in number of sites) between two correlated spins, as obtained in Ref.[\onlinecite{Antal}]. %In addition, this result demonstrates the success of the iTEBD method in the critical regime. 
The specific power-law is considered to govern the critical region of non-equilibrium steady-state models in general \cite{Torre}.
The correlation function is written as:
\begin{equation}
		\langle \sigma^z_n \sigma^z_{n+R} \rangle \sim \frac{Q(h,L)}{\sqrt{R}}.\cos{(k R)}
	\end{equation}
The value of $Q(h,L)$ can be calculated exactly in the limit $L \rightarrow \infty$, as the limit of the amplitude of the same correlations of the XY-model in 1D \cite{Baruch}, with the result $Q(h, L=\infty) = e^{1/2} 2^{-4/3}A^{-6} \simeq 0.147$, where $A  \simeq  1.282$ is the Glaisher's constant. For different values of $L$, Ref.~[\onlinecite{Antal}] approximated $Q(h,L)$ away from the phase boundaries with:
\begin{equation}
Q(h,L) \simeq Q(h,\infty) \left(\frac{L^2-h^2}{L^2-J^2}\right)^{1/4}
\end{equation}
while, close to the boundary $h=J$, $Q(h=J,L) \simeq Q(h, \infty) (\frac{L^2}{L^2-J^2})^{1/8}$.
The wavenumber $k$ of the spatial dependence of the correlations is independent of the magnetic field \cite{Antal} and as $L \rightarrow \infty$ it is given by:
\begin{equation}
	\label{eq:wavenumber}
	k=\arccos(1/L).
\end{equation}

%Using least squares fitting, we find an approximate expression for the damped sinusoidal correlation function,
%	\begin{equation}
%		\langle \sigma^z_n \sigma^z_{n+R} \rangle \sim \frac{Q(h,L)}{\sqrt{R}} \cos{(k R)}
%	\end{equation}
%The amplitude decreases strongly as the phase boundary is approached $h\rightarrow 2J$, more details on the analysis of $Q(h,L)$ have been discussed in Ref.\onlinecite{Antal}.

%\begin{figure}[htp]
%	{\includegraphics[width=0.48\textwidth]{graphs/TFIM_wc/Q_vs_L}}
%	\vspace{-0.3 cm}
%	\caption{\label{fig:Q_vs_L} The amplitude of oscillatory correlations as a function of L.}
%\end{figure}

\begin{table}
  \caption{\label{tab:period} Numerical data showing how the wavelength $\cal{R}$ and wavenumber $k$ vary with current $L$}
  \begin{tabularx}{0.5\textwidth}{XXXXX}
    \hline\hline
    $L$ & $\cal{R}$ & $k$\\
    \hline
   	1.2 & 10.31 & 0.61\\
	1.4 & 7.67 & 0.82\\
	1.6 & 7.29 & 0.86\\
	1.8 & 6.10 & 1.03\\
	2.0 & 5.98 & 1.05\\
    \hline\hline
  \end{tabularx}
\end{table}

%Eq. (\ref{eq:wavenumber}) is exact in the limit $\Lambda \rightarrow 0$. 

Numerically, the correlations are computed using the iTEBD algorithm and the critical correlations in the energy current-carrying region has been verified. Using Fast Fourier Transform (FFT), the peak of the oscillations in space, determines the wavelength ${\cal{R}} = 2 \pi/k$. This is shown in Fig.~\ref{fig:FMIsing}).Table I lists results for current, wavelength and wavenumber. A least squares fit of the data indicates that Eq. (\ref{eq:wavenumber}) is in general reliable and we find that:
\begin{equation}
	k= (1.01\pm 0.06)\arccos[(0.99\pm 0.05)/L].
\end{equation}

\subsection{Entanglement scaling and central charge}
The universal entanglement properties of the TFIM are accessible within conformal field theory \cite{Calabrese}. A universal formula for the entanglement entropy \[ S \sim \frac{c}{6} \log{\left(\xi/a\right)}\] depends on the central charge $c$ and the correlation length $\xi$ ($a$ is the lattice spacing). In this expression for the entanglement entropy the prefactor is 1/6 (and not 1/3), due to the fact that we consider an infinite DMRG algorithm in the calculations, therefore there is only one contact point between the two parts of the spin chain, to be taken into account. This is contrasted to the case of a chain with periodic boundary condition.

To study the properties of the system, at a quantum critical point, where the correlation length is infinite, within the MPS framework, we can perform an entanglement scaling, as $\chi \rightarrow \infty$ \cite{pollmann09,tagliacozzo}, in the spirit of the work of Ref. [\onlinecite{White93}]. 
More specifically, the entanglement entropy is calculated from the Schmidt decompositions singular values using the von Neumann entropy
\begin{equation}
		S = -\sum_{a_n}\left(\lambda^{[0]}_{a_n}\right)^2 \ln{\left(\lambda^{[0]}_{a_n}\right)^2}= -\sum_{a_n}\left(\lambda^{[1]}_{a_n}\right)^2 \ln{\left(\lambda^{[1]}_{a_n}\right)^2}
\end{equation}
The values are identical for either half of the chain (odd or even bond) and the bipartite splitting of the MPS has entropy that is maximally entangled at $\log{\chi}$.

The information on the critical properties of the system can be obtained by employing finite entanglement scaling. In a critical chain the scaling equation
\begin{equation}
	\xi \propto \chi^\kappa
\end{equation}
holds, where $\chi$ is the number of states kept and $\kappa$ is a constant  that depends only on the central charge of the model~\cite{pollmann09}.
This method to extract the central charge is followed in the next sections as well and more details are presented in Appendix A.

\begin{figure}[htb]
  {\label{}\includegraphics[width=8cm]{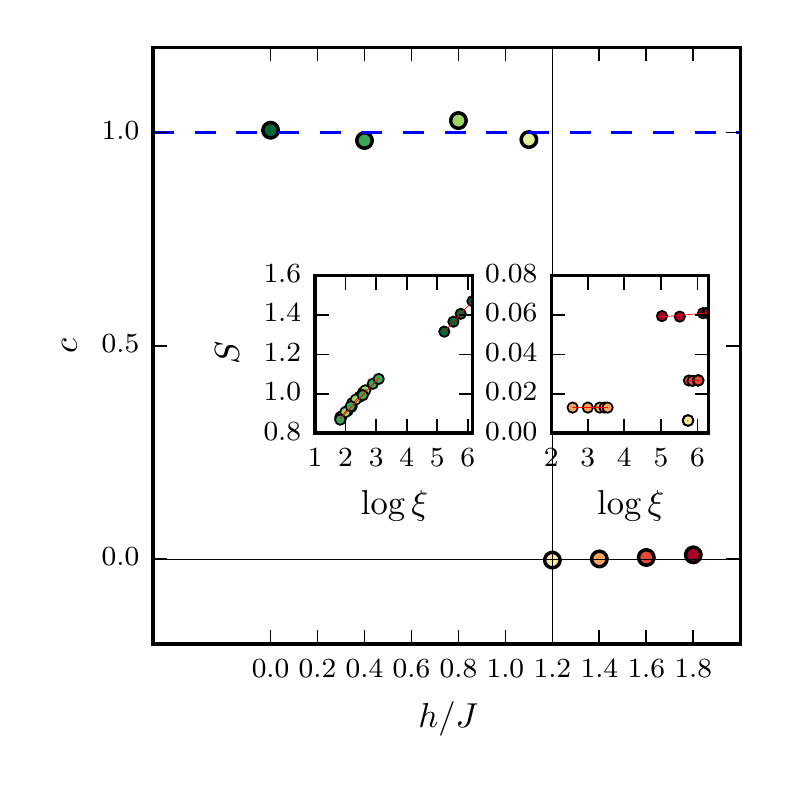}}
  \centering
  \caption{Central charge as a function of the transverse field $h$ in the current-carrying TFIM at $L=1.2J$. In the inset we show an example of how the central charge $c=1$ is determined in the current-carrying TFIM (at $h=0.5J$ and $L=1.2J$) from the calculation of $S=\frac{c}{6} \log(\xi)$.}
  \label{fig:entanglement2}
\end{figure}

From the calculation of the entanglement entropy, we extract the central charge $c$ in the current-carrying region which is critical with an expected central charge $c=1$. A change in $c$ naturally occurs at the boundary to the gapped region where $c=0$.
For example at fixed current $L=1.2$ and non-critical field $h=0.5J$ the system sits in the current region. The typical scaling behavior of the entanglement entropy is presented in in the inset of Fig.~\ref{fig:entanglement2}).

The system remains critical in the entire current carrying region. The transverse field dependence of the scaling is shown in Fig.~\ref{fig:entanglement2}. At a small growing perturbation in $h\ll L$ is observed a denser set of points on the left-hand side of Fig.~\ref{fig:entanglement2}, and shows a central charge $c=1$. Further increase of the magnetic field strength into the non-critical paramagnetic (PM) region hits the boundary at $h=L$ where the central charge is $c=0$. 
Multiple points around $L=1.2J$ show numerical noise which is due to instabilities near the transition. 

\section{TFIM with a perturbation that breaks integrability.}
Integrability breaking can be achieved by introducing certain perturbation terms in the TFIM. One way that this can be achieved in the TFIM is through the introduction of an interaction $D \sigma_i^x \sigma_{i+1}^x$ longitudinal to the magnetic field direction $x$ and transverse to the original spin ordering interaction $z$. 
The Hamiltonian reads
\begin{equation}
	\label{}
    	H_{non}=-\sum_{i=1}^N J  \sigma_i^z \sigma_{i+1}^z + h  \sigma_i^x + D \sigma_i^x \sigma_{i+1}^x.
\end{equation}

We choose to work with this Hamiltonian, as it is to large extent unexplored compared to other ways to break the integrability of the system. 
%The model as $D\rightarrow0$ approaches the TFIM. It is expected that creating a small perturbation by gradually increasing $D$ will move the system out of integrability and into an area of new physics.
%Before an attempt at a numerical simulation, some discussion of known solutions to the model is informative. 
If the spin couplings have equivalent interaction strength $J$=$D$=1, the Hamiltonian is the isotropic XY model with an integrability breaking $x$-directed magnetic field. Additionally, in the absence of magnetic field $h$=0, 
%if the interaction strength $D$ is no longer constrained to equal $J$, i.e. it holds any value 
and $D\geq J$,  the anisotropic XY model emerges This has a known exact solution, with energy dispersion relation 
\begin{equation}
	%\epsilon^{non}_{h=0,k}=J \left(1-[1-(1-2(D/J))^2]\sin^2(ka)\right)^{1/2}.
	\epsilon^{non}_{h=0, k}=J\left(1-\left[1-\left(\frac{1+D/J}{1-D/J}\right)^2\right]\sin^2(k)\right)^{1/2}
\end{equation}
As a consequence, there are two limits where the system is integrable with known critical entanglement properties.

\subsection{Spin-spin correlations}
\begin{figure}[!h]
	{\includegraphics[width=8cm]{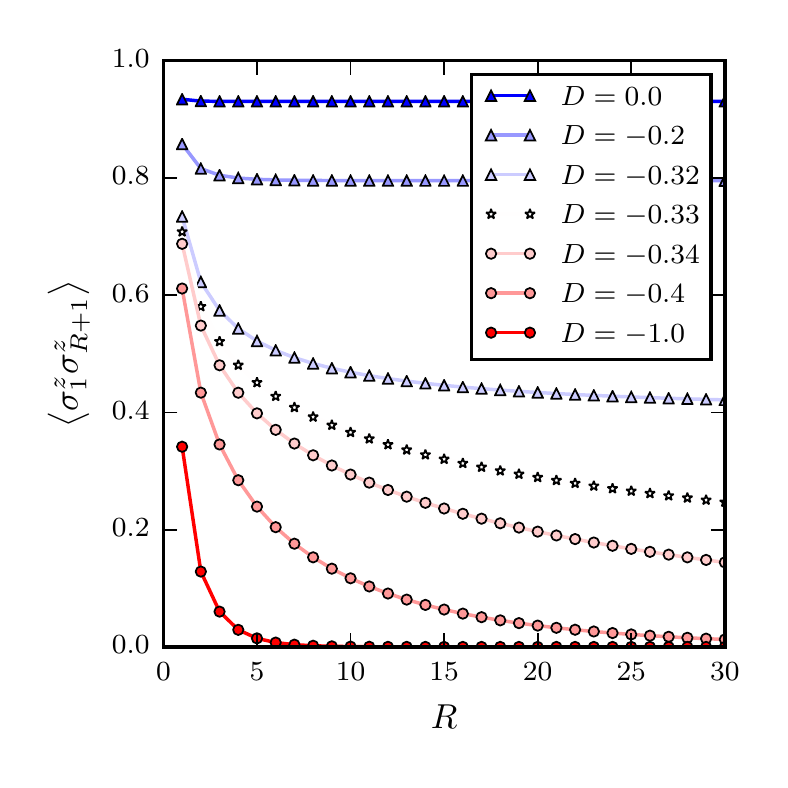}}%
	\caption{Correlations of the non-integrable TFIM at $h$=0.5 with integrability breaking parameter $D$.}
	\label{fig:NI_correlations_D}
\end{figure}

The system passes through a critical point as $D$ increases, the $zz$-correlations are presented in Fig.~\ref{fig:NI_correlations_D}. At low values of $D$ there is long-range order 
until the critical point is reached at $D_c$, correlations with power law decay emerge. The non-integrable term acts similarly to a transverse field, and at $D_c$, the system experiences a phase transition from a  $Z_2$ symmetry breaking FM to PM phase.

\subsection{Entanglement scaling and central charge}
There are two known critical points at $h=J, D=0$ and $h=0,D=J$; the TFIM and the isotropic XY (or XX) limits, with central charges $0.5$ and $1$, respectively~\cite{Vidal2}.
A critical line connects these two points.
%To account for an expected phase transition when $D$ is varied, a critical line with nonzero central charge should connect these two points. Away from this line the model is non-critical with central charge $0$.
The full phase diagram is found numerically using a combination of the iDMRG algorithm and finite entanglement scaling as shown in Fig.~\ref{fig:D_vs_h}. As demonstrated in Appendix~A, the critical line that connects the two known critical points is located by observing a peak in the effective correlation length $\xi$ as it increases with $\chi$. The entanglement scaling along the line maintains $c=0.5$ until the XY point is reached, where $c=1$. This is in complete analogy with the anisotropic XY model which only at the point of isotropy (XX model) the universality class is of the free boson with central charge c=1 but away from that point, in the critical region, the universality class is free fermions with c=1/2 \cite{Latorre2}. We see the same physics by perturbing the system differently in this case. 
%The rest of the digram shows no scaling as the parameters tune the system away from criticality.

\begin{figure}[!h]
	\makebox[\textwidth][l]{\includegraphics[width=0.50\textwidth]{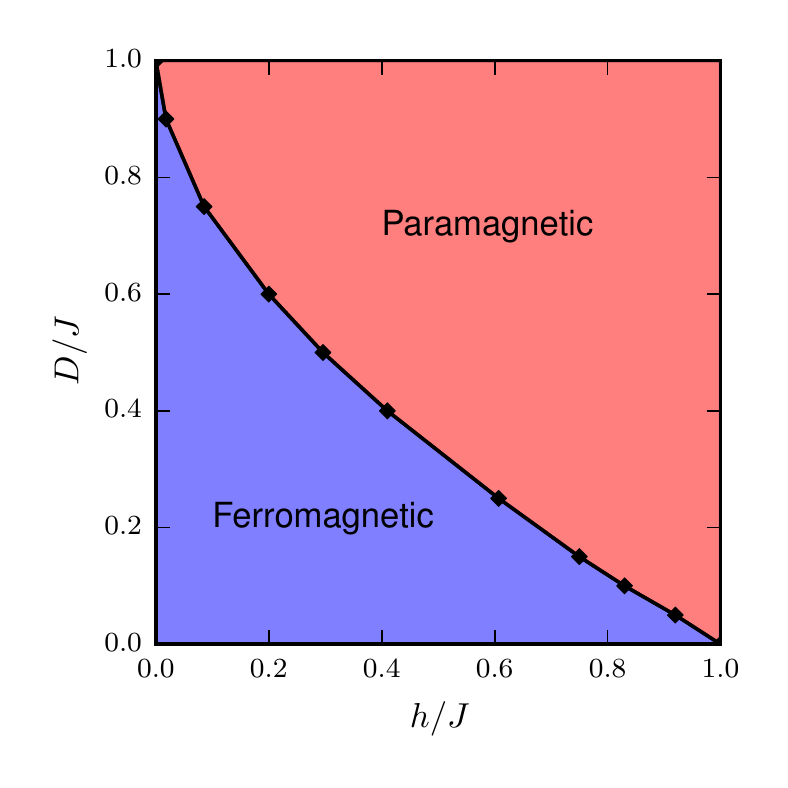}}%
	\caption{Phase diagram of the TFIM with integrability breaking term as function of the transverse field $h$ and the nearest-neighbor interactions.}
	\label{fig:D_vs_h}
\end{figure}

\section{TFIM with a periodically weakened bond.}
\subsection{Diagonalization of Hamiltonian}
So far, the two variations of the TFIM
%(\ref{Ising_wc}) 
describe homogeneous two-site interactions across the length of the spin chain. If we consider weakening the coupling at every even site, the Hamiltonian can be written as:  
\begin{equation}
		\label{Ising_wc}
    	H_{wb} =-\sum_{i=1}^N (J \sigma_{i}^z \sigma_{i+1}^z  + h \sigma_i^x)-\sum_{j=1}^{[N/2]} (J' -J) \sigma_{2j}^z \sigma_{2j+1}^z
\end{equation}

This is equivalent to defining two Majorana fermion populations with two independent particle number operators. The Hamiltonian can be diagonalised with a four-dimensional Bogoliubov transformation \cite{Jafari} with the dispersion relation:

\begin{widetext}	
\begin{equation}
	\label{exact_weak_bond}
	\epsilon^{wb}_k = \left[ 2J^2+2J'^2+4h^2  + 2\sqrt{(J^2-J'^2)^2+4h^4(J^2+J'^2) +8JJ'h^2\cos{ka}}\right]^{1/2}
\end{equation}
\end{widetext}

\begin{figure}[htp]
  \begin{center}
    {\label{}\includegraphics[width=8.5cm]{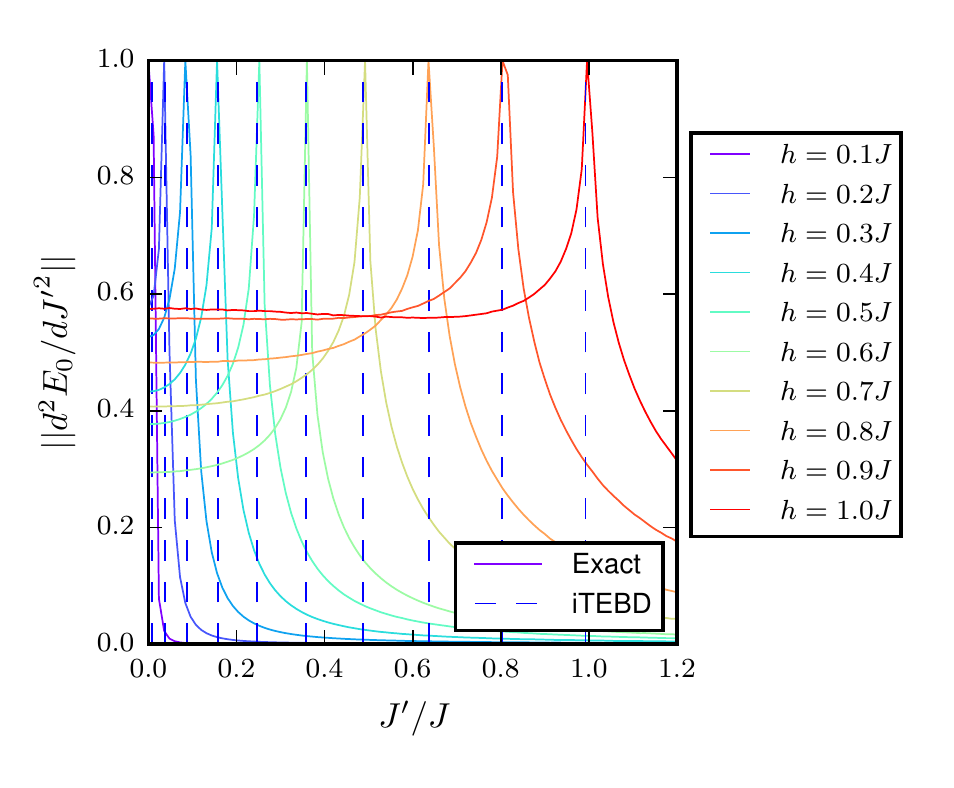}}
  \end{center}
  \caption{The second derivative of the weak bond Ising energy dispersion relation with respect to $J'$. The peaks are normalized for clarity since their magnitude decays with increasing field. The divergence indicates quantum critical points as $d^2 E_0/dJ'^2 \to \infty$. The divergence is also calculated numerically from the ground state energy using iTEBD and the critical points are depicted as vertical dashed lines.}
  \label{fig:exact_weak_bond}
\end{figure}

Integrating  this equation over $k$ for $N\rightarrow \infty$, the exact ground state energy is calculated. The second derivative of the energy with respect to $J'$ diverges at sufficiently weak $J'$, Fig.~\ref{fig:exact_weak_bond}. A second-order phase transition occurs as the system passes through this point. This becomes clear from the form of the order parameter, which changes continuously to zero upon approaching the transition bond strength. The behavior is universal and identical to the situation where the system passes through the critical point by varying the magnetic field in the TFIM.

\subsection{Spin-spin correlations}
\begin{figure}[htp]
  \begin{center}
    {\label{}\includegraphics[width=9cm]{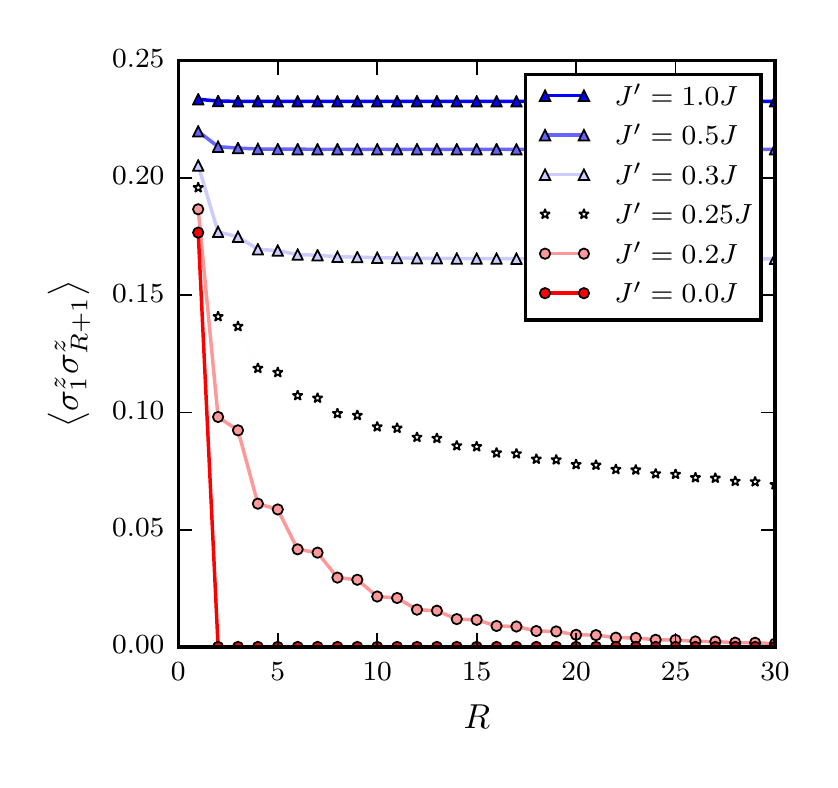}}
  \end{center}
  \caption{Correlations for the TFIM with weakened bond at $h$=0.5.}
  \label{fig:correl_weak_bond}
\end{figure}

Fig.~\ref{fig:correl_weak_bond} shows the correlations as every even bond is weakened through the transition point at $J'_c=0.25J$ when $h=0.5J$. This coincides with the peak in Fig.~\ref{fig:exact_weak_bond}. Long-range order is maintained until $J'$ is tuned to $J_c$ and power-law decay appears. On the other side of the transition $J_c<0.25J$ the region experiences exponential decay. When the bond strength is completely diminished $J'=0$ and the chain is isolated into pairs. The correlations are instantaneously cut-off and the chain ceases to correlate beyond the first neighbour.
 
Short-range interactions are modified from the usual equivalent bond strength TFIM. The odd bond preserves a modulated pairing between spins experiencing the full strength $J$. Correlations show an almost sawtooth decay instead of a smooth curve.

\subsection{Entanglement scaling and central charge}
The entanglement scaling properties are investigated at the divergent points in Fig.~\ref{fig:exact_weak_bond}. The critical nature at $(J'_c/J) =(h/J)^2$ shares universal properties with the TFIM. This suggests the entanglement scaling should exhibit similar behavior. 

\begin{figure}[htp]
 \begin{center}
    {\label{}\includegraphics[]{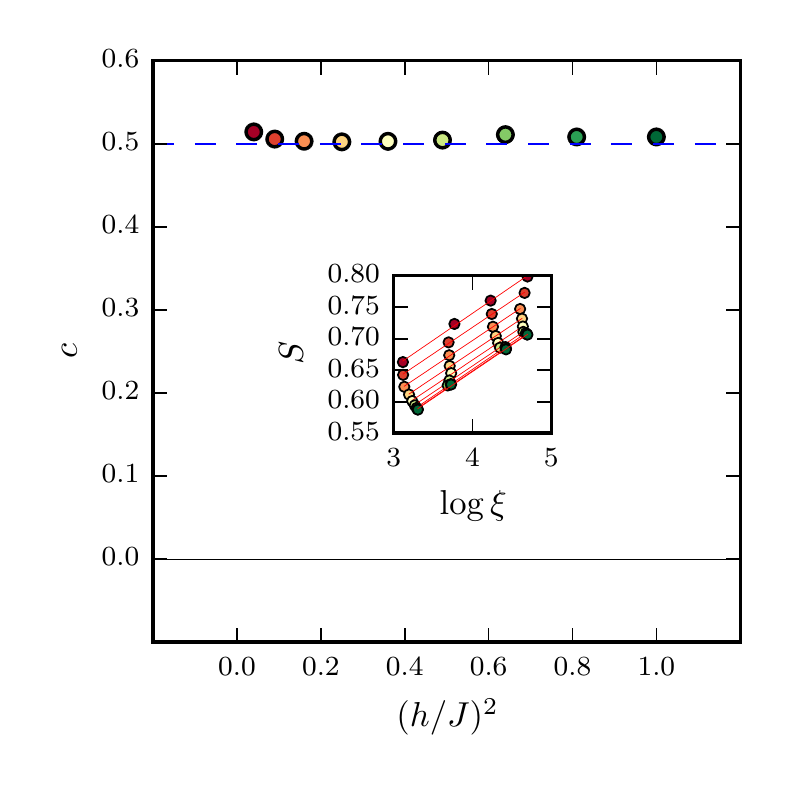}}
  \end{center}
  \caption{Central charge as a function of $(h/J)^2$ in the weakened bond transverse field Ising model at critical $J'$. Inset is the entanglement scaling where the central charge of $c$=0.5 is extracted from.}
  \label{fig:central_charge_bond_defect}
\end{figure}

Fig.~\ref{fig:central_charge_bond_defect} presents a plot of the central charge at the critical points. The central charge is identical to the TFIM. Otherwise when $J'/J \neq (h/J)^2$ the system is tuned away from criticality, the scaling disappears and the central charge is zero.

\subsection{In the presence of energy current}
\begin{figure}[htp]
  \begin{center}
    {\label{}\includegraphics[width=0.48\textwidth]{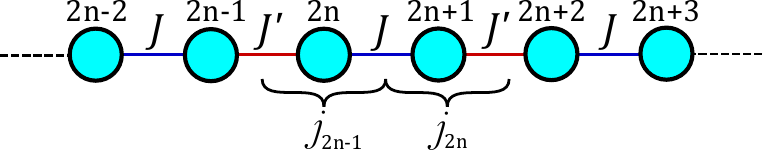}}
  \end{center}
    \caption{Illustration of the sites, the corresponding interactions, the energy currents and the bonds that are involved. Due to the symmetry of the problem, the energy current $j_{2n}$ and $j_{2n+1}$ need to be considered in the continuity equation.}
    %Introducing current ($L$=2) to the TFIM with bond impurity at finite field strength $h=0.5$. $J'>0.5$ correlations are rigid and consistent with \ref{Ising_wc}, at sufficient weakening the model undergoes two changes, the period decreases and amplitude decay is mitigated until disorder is reached at $J'=0$.}
  \label{fig:FM_impurity_current}
\end{figure}

In the situation where an energy current is introduced in the system, in a similar way as in the homogeneous TFIM, then it can be proved that there is no energy current conservation. The system then is not in a steady state. To show this we shall take the energy current over two consecutive bonds, as schematically seen in Fig.~\ref{fig:FM_impurity_current} and find its time derivative. The energy current density over two sites sites $2n, 2n+1$ that involves three bonds as indicated in Fig.~\ref{fig:FM_impurity_current}  reads:

\begin{widetext}
\begin{equation}
j_{2n} + j_{2n+1} = J' h \sigma_{2n-1}^z \sigma_{2n}^y - J h \sigma_{2n}^y \sigma_{2n+1}^z 
+ J h \sigma_{2n}^z \sigma_{2n+1}^y -J'h\sigma_{2n+1}^y \sigma_{2n+2}^z 
\end{equation}
\end{widetext}

The continuity equation then takes the form:

\begin{widetext}
\begin{eqnarray}
\nonumber
\partial_t (j_{2n} + j_{2n+1}) = - i \left[ H, j_{2n} + j_{2n+1} \right] =&& 2 h (J'^2 - J^2) (\sigma_{2n}^x - \sigma_{2n+1}^x) +2J'h^2 \left[ \left(\sigma_{2n-1}^y \sigma_{2n}^y -  \sigma_{2n-1}^z \sigma_{2n}^z\right) \right.\\
&-& \left. \left( \sigma_{2n+1}^y \sigma_{2n+2}^y -  \sigma_{2n+1}^z \sigma_{2n+2}^z \right) \right]
\end{eqnarray} 
\end{widetext}

The form of the above equation indicates that upon summation over all sites, the last term on the right hand side, proportional to $J'h^2$, will vanish except from the boundaries but the first term on the right hand side will only vanish if $J'=J$. Thus there is no energy current conservation in the system, contrary to the homogeneous case. 

%In the interesting situation of an energy current through the modulated chain there is an interplay between the defected bond $J'$ and the energy current (details can be found in Appendix B). As it is expected, as the critical point is passed, then the system is taken out of the region with current flow and into the region of disorder. The correlations are presented in Fig. \ref{fig:FM_impurity_current}. For $J'>h$, the correlations are similar to the homogeneous case in Fig.~\ref{fig:FMIsing}, the strength of the weakened bond is sufficient to maintain the chains' original correlations. 

%\begin{equation}
%	J^E_{wb}= -\sum_{j=1}^{N/2} \frac{h}{2}(J'\sigma_{2j-1}^z \sigma_{2j}^y - J\sigma_{2j}^y \sigma_{2j+1}^z)
%\end{equation}
%
%\begin{equation}
%	\eta^{wb}_k= \epsilon^{wb}_k + L(1+J'/J)\sin(ka)
%\end{equation}
%

\subsection{Generalization to $n$-site period}
\begin{figure}[htp]
  \begin{center}
    {\label{}\includegraphics[width=8cm]{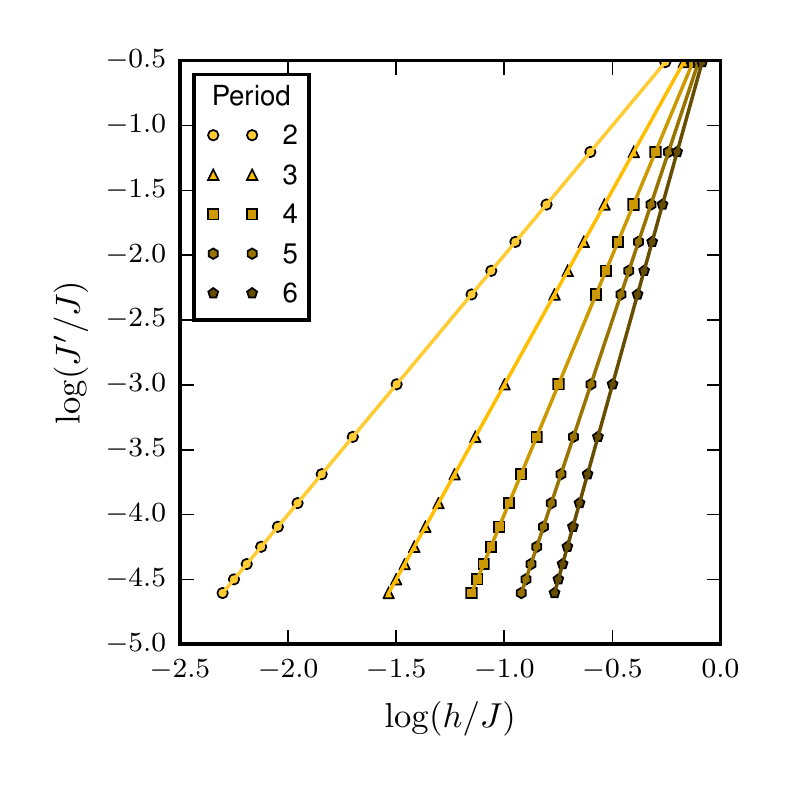}}
  \end{center}
  \caption{Critical boundary for transverse Ising model with weakened bond over $n$-sites, $(J'_{c}/J)=(h/J)^n$. Second-order phase transition occurs crossing the line from above (ordered to disordered). }
  %(b) Schematic diagram introducing impurity on third axis. Average current $\left<j\right>=0$ for $L\leq J$, having no influence on the boundary.}
  \label{fig:boundary}
\end{figure}

The previous discussion considered the weakening of every second bond. The generalisation to a
periodically weakened bond of a period $n$-sites reveals a relationship for the critical $J'$ that scales with the transverse field as
%for an impurity separated periodically by $n$-sites is found to be 
$J'_{c}/J=(h/J)^n$ . Figure~\ref{fig:boundary} (a) shows the critical lines for $n$ from 2 to 6.

The physics behind  the above result, is that as the distance between the weakened bonds becomes larger, then lower values of $J'$ are required to produce the same effect and, in the limit of very large $n$ then $J'_{c}$ approaches $0$ which indicates that it effectively requires cutting the chain in order to produce the same result. The analytical treatment and proof of this relation is beyond the scope of the present work and will be presented elsewhere as it involves the non-trivial diagonalisation of a chain of $n$ spins with periodically modulated couplings and periodic boundary conditions \cite{Feldman}. The case of a single site with transverse field in an $n$-periodic chain has been investigated analytically, e.g., in Ref.~[\onlinecite{Genovese}]; the treatment there is simplified by the fact that the defect is on the site rather than on the bond.

%A new boundary is present if a new $J'$ axis is introduced. This is identified as a new disordered region below the ramp-like boundary in Fig.\ref{fig:boundary} (b). 

\section{Discussion}

In this work we present a detailed study of entanglement entropy and critical correlations of the one-dimensional TFIM in three cases using mainly numerical calculations. We have revisited the TFIM with an energy current confirming the phase diagram and correlations \cite{Antal} and the entanglement properties where we found that the central charge $c$ takes the value 1 instead of 1/2 which was also analytically calculated \cite{Kadar}. Similar change of the central charge was seen in the entanglement scaling for the XX chain in the current carrying region \cite{Eisler2}. The reason is that the Fermi sea is now doubled. For symmetrically arranged multiple Fermi seas it has been shown \cite{Keating} that the entanglement is proportional to the number of the Fermi seas.  This has been generalised to arbitrary arrangement of the Fermi seas and to cases without particle-number conservation.
It is worth mentioning that in Ref.~[\onlinecite{Kadar}], it was pointed out that there exists a duality relation between the TI chain with current and XX chain with current.

Then as a natural continuation, we added an integrability-breaking term (an interaction of the form $\sigma^z \sigma^z$) and studied the entanglement properties. We found a critical line that connects two known critical points; the pure TFIM and the isotropic XY (or XX). Away from the point with XX symmetry, the central charge is c=1/2 along the critical line.

Subsequently by taking the TFIM and altering every other bond (from $J$ to $J'$), we analyzed the energy dispersion relation and found the critical points. By adding a current we showed that there is no energy current conservation, which is only recovered if $J=J'$. We generalized the model to one with $n$-site periodicity and found a critical $J'$ that scales as $J'_{c}/J=(h/J)^n$, where $n$ is the distance of two adjacent weakened bonds. This demands further analytical treatment to be reported elsewhere. It should be noted that there have been studies of isolated impurities of quantum spin chains, via bosonization \cite{Eggert} and most recently using numerical techniques \cite{Biella}.

Overall, taking a TFI chain as a basic model with scientific and also practical (such as in quantum information) consequences, we provide new results in different variants by using matrix-product state based methods. The entanglement properties of quantum XY spin chains of arbitrary length has been investigated by using the notion of global measure \cite{Wei1, Wei2}. In earlier works it was found that the field derivative of the entanglement density
becomes singular along the critical line. The form of this singularity is dictated by the universality class
which controls the quantum phase transition. Moreover, it was pointed out that there is a deeper connection between  the global entanglement and the correlations among
quantum fluctuations. This is another direction to be clarified.

\section*{Acknowledgements}
We would like to thank Sasha Balanov, Claudio Castelnovo, Fabian Essler, Alex Zagoskin for useful discussions and Mike Hemsley for early collaborations on related problems. The work has been supported partly by EPSRC through grant EP/M006581/1 (JJB) and by Research Unit FOR 1807 through grants no. PO 1370/2-1 (FP).

\section*{Appendix A: Finite entanglement scaling}
In infinite matrix product state methods the length of the system is no longer a limiting factor. The finite size scaling is traded for finite entanglement of the state  with the bond dimension $\chi$ used as the new scaling resource, which now limits the representation of the state. The universal formula for the entanglement entropy is
\begin{equation}
	S \sim \frac{c}{6} \log{\left(\frac{\xi}{a}\right)}
\end{equation}
and depends on the correlation length $\xi$ calculated from the two dominant eigenvalues of the quantum transfer matrix
\begin{equation}
 	\xi = 1/\ln(\lambda_1/|\lambda_2|) = -1/\ln|\lambda_2|
\end{equation}
As the MPS has been built in canonical form with orthonormal left and right eigenvectors, the first eigenvalue in the transfer matrix is 1, thus we need to calculate the second eigenvalue. The correlation length scales inversely with energy gap until it diverges at the critical point when the gap vanishes. 
%The constant $a$ is not important as the gradient of the equation is the only value of concern used to find the central charge.

\begin{figure}[!h]
	\makebox[\textwidth][l]{\includegraphics[width=0.50\textwidth]{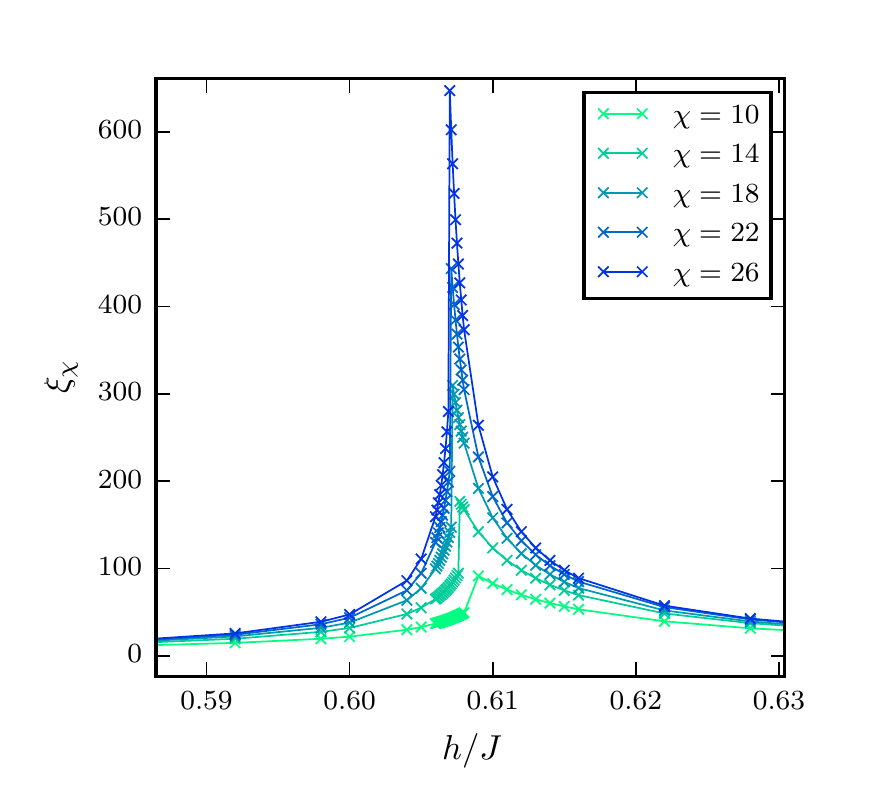}}%
	\caption{Effective correlation length at $D=0.25$, critical $h\approx 0.607$.}
	\label{fig:FES}
\end{figure}

As explained in Sec. II, the entanglement entropy is calculated from the Schmidt decomposition's singular values using the von Neumann entropy,
%\begin{equation}
%		S = -\sum_{a_n}\left(\lambda^{[0]}_{a_n}\right)^2 \ln{\left(\lambda^{[0]}_{a_n}\right)^2}= -\sum_{a_n}\left(\lambda^{[1]}_{a_n}\right)^2 \ln{\left(\lambda^{[1]}_{a_n}\right)^2}
%\end{equation}
%The values are identical for either half of the chain (odd or even bond) and the bipartite splitting of the MPS has entropy that is maximally entangled at $\log{\chi}$.
followed by a finite entanglement scaling. In a critical regime $\xi \propto \chi^\kappa$.
Although the perfect description of the state at a critical point requires $\chi$ to diverge to infinity, the fact that $\chi$ is finite leads to an effective finite correlation length $\xi_{\chi}$ at the critical point. The phase transition is seen at the peak in a curve as the control parameter passes through the critical point. Depending on the number of states retained the peak can be broad for low values of $\chi$ and give an  less accurate representation of the exact critical point. For larger values of $\chi$ the peak becomes sharper and more reliable. This is seen in Fig.~\ref{fig:FES} which is an example of how the critical points in Section III were calculated.

\begin{figure}[!h]
	\makebox[\textwidth][l]{\includegraphics[width=0.50\textwidth]{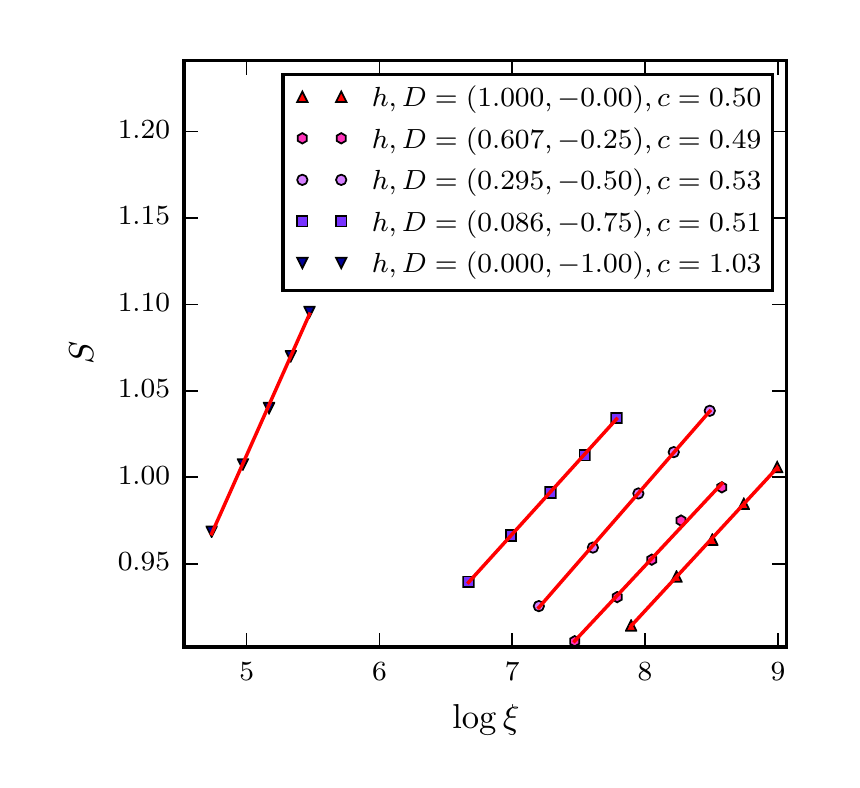}}%
	\caption{Central charge along the non-integrable critical line.}
	\label{fig:FES_non_int}
\end{figure}

Furthermore, in Fig.\ref{fig:FES_non_int} we illustrate by using examples, how the central charge is determined from the scaling of entanglement entropy.

\section*{Appendix B: Energy current derivation}
%By applying a form of energy flow, inducing a non-equilibrium state, the behavior of systems under flux can be investigated.
Application of energy current creates a non-equilibrium \emph{steady-state} if the time derivative of the total energy current is zero. In a steady-state energy flows through the system at a constant rate continuously, without any impulses causing discontinuous energy transfer. The energy is conserved and thus characterised by a conservation law. If the Hamiltonian of the system can be written as sum of terms that depend on two sites: $H=h_{1,2}+\cdots+h_{i,i+1}+\cdots+h_{N,N+1}$ and $[h_{i,i+1},h_{i-1,i}]\neq 0$ then the conservation of energy is expressed in the form of a continuity equation~\cite{Zotos}
\begin{equation}
	 \frac{\partial h_{i,i+1}(t)}{\partial t}
	   +\Delta j_i = 0
\end{equation}
$\Delta j_i=j_{i+1}-j_{i}$ is the discrete divergence of the current.

Applying a unitary transformation (working in the Heisenberg picture) \[h_{i,i+1}(t)=U h_{i,i+1}U^{-1}=e^{iHt}h_{i,i+1}e^{-iHt}\] the time derivative is:
%\begin{align}
%	\frac{\partial h_{i,i+1}(t)}{\partial t} &= U h_{i,i+1}\frac{\partial (U^{-1})}{\partial t}+U\frac{\partial (h_{i,i+1})}{\partial t}U^{-1} \nonumber \\
%	&\quad+\frac{\partial (U )}{\partial t}h_{i,i+1}U^{-1} \\
%	&= U h_{i,i+1}\left(-i H\right) U^{-1} + \left(i H\right) U h_{i,i+1}U^{-1} \\
%	&= i[H,h_{i,i+1}(t)]
%\end{align}
\begin{eqnarray}
\nonumber
\frac{\partial h_{i,i+1}(t)}{\partial t} &=& i[H,h_{i,i+1}(t)]\\
\nonumber
&=& i [h_{i-1,i},h_{i,i+1}(t)]-[h_{i,i+1}(t),h_{i+1,i+2}]
\end{eqnarray}

\noindent where we have used the relations $dU/dt=iHU$ and $[H,U]=[H,U^{-1}]=0$.
%Further, writing the Hamiltonian as $H=h_{1,2}+\cdots+h_{i,i+1}+\cdots+h_{N,N+1}$ and $[h_{i,i+1},h_{i-1,i}]\neq 0$, we obtain
%\begin{align}
%	i[H,h_{i,i+1}(t)] &= i H h_{i,i+1}(t) - i h_{i,i+1}(t) H  \\
%	&= i(h_{i-1,i} \; h_{i,i+1}(t) +h_{i+1,i+2} \; h_{i,i+1}(t)\nonumber \\
%	&\quad -i\left(h_{i,i+1}(t) \; h_{i-1,i}+h_{i,i+1}(t) \; h_{i+1,i+2}\right) \\
%	&= i[h_{i-1,i},h_{i,i+1}(t)]-i[h_{i,i+1}(t),h_{i+1,i+2}]
%\end{align}
%\begin{equation}
%	[H,h_{i,i+1}(t)] = [h_{i-1,i},h_{i,i+1}(t)]-[h_{i,i+1}(t),h_{i+1,i+2}]
%\end{equation}
As the two commutation relations on the right hand side depend on the previous $i-1$ and next $i+1$ sites, they can be identified as the local current operators $j_i$ and $j_{i+1}$ respectively, meaning the continuity equation is
\begin{equation}
	\frac{\partial h_{i,i+1}(t)}{\partial t}=j_i(t)-j_{i+1}(t)
\end{equation}
\noindent with $j_i(t) = i[h_{i-1,i},h_{i,i+1}(t)]$.
%therefore we have explicitly calculated the two local current operators in $\Delta j_i$. 
We now apply the above for the two cases of TFIM we have considered, finding:

\noindent\textit{1. Energy current in TFIM}
%\begin{align}
%	j^E_i&=-i[J \sigma_{i-1}^z \sigma_{i}^z + h (\sigma_{i-1}^x+\sigma_{i}^x),J \sigma_{i}^z \sigma_{i+1}^z + h (\sigma_i^x+\sigma_{i+1}^x)] \nonumber\\
%	&=-i\big(J^2[\sigma_{i-1}^z \sigma_{i}^z, \sigma_{i}^z \sigma_{i+1}^z]+Jh[ \sigma_{i-1}^z \sigma_{i}^z,(\sigma_i^x+\sigma_{i+1}^x)] \nonumber\\
%	&\hspace{0.5cm}+Jh[(\sigma_{i-1}^x+\sigma_i^x),\sigma_{i}^z \sigma_{i+1}^z]+h^2[(\sigma_{i-1}^x+\sigma_i^x),(\sigma_i^x+\sigma_{i+1}^x)]\big) \nonumber\\
%	&=-i\big(Jh(\sigma_{i-1}^z[\sigma_i^z,\sigma_i^x]+[\sigma_i^x,\sigma_i^z]\sigma_{i+1}^z)\big) \nonumber\\
%	&= Jh(\sigma_{i-1}^z \sigma_{i}^y - \sigma_i^y \sigma_{i+1}^z)
%\end{align}
\begin{equation}
	j^E_i = Jh(\sigma_{i-1}^z \sigma_{i}^y - \sigma_i^y \sigma_{i+1}^z)
\end{equation}

\noindent\textit{2. Energy current in TFIM with alternating bond strength}
%\begin{align}
%	j^E_i&=-i[J' \sigma_{i-1}^z \sigma_{i}^z + h (\sigma_{i-1}^x+\sigma_{i}^x),J \sigma_{i}^z \sigma_{i+1}^z + h (\sigma_i^x+\sigma_{i+1}^x)] \nonumber\\
%	&=-i\big(J'J[\sigma_{i-1}^z \sigma_{i}^z, \sigma_{i}^z \sigma_{i+1}^z]+J'h[ \sigma_{i-1}^z \sigma_{i}^z,(\sigma_i^x+\sigma_{i+1}^x)] \nonumber\\
%	&\hspace{0.5cm}+Jh[(\sigma_{i-1}^x+\sigma_i^x),\sigma_{i}^z \sigma_{i+1}^z]+h^2[(\sigma_{i-1}^x+\sigma_i^x),(\sigma_i^x+\sigma_{i+1}^x)]\big) \nonumber\\
%	&=-i\big(J'h\sigma_{i-1}^z[\sigma_i^z,\sigma_i^x]+Jh[\sigma_i^x,\sigma_i^z]\sigma_{i+1}^z\big) \nonumber\\
%	&= h(J'\sigma_{i-1}^z \sigma_{i}^y - J\sigma_i^y \sigma_{i+1}^z)
%\end{align}
\begin{equation}
	j^E_i	= h(J'\sigma_{i-1}^z \sigma_{i}^y - J\sigma_i^y \sigma_{i+1}^z)
\end{equation}

\noindent if the strength of the interaction (bond) is $J' $ between sites ${i-1,i}$ and $J$ between ${i, i+1}$. Note that in this case, to derive the continuity equation we need to take into account the flow of energy density passing through two adjacent sites, due to the doubling of the unit cell.

%\\
%\noindent\textit{2. Energy current in TFIM with an additional integrability-breaking term}
%\begin{align}
%	j^{E_{non}}_i&=-i[J \sigma_{i-1}^z \sigma_{i}^z + D \sigma_{i-1}^x \sigma_{i}^x + h (\sigma_{i-1}^x+\sigma_{i}^x), \nonumber \\
%	& \quad \qquad J \sigma_{i}^z \sigma_{i+1}^z + D \sigma_{i}^x \sigma_{i+1}^x + h (\sigma_i^x+\sigma_{i+1}^x)] \nonumber\\
%	&=-i\big(J^2[\sigma_{i-1}^z \sigma_{i}^z, \sigma_{i}^z \sigma_{i+1}^z]+JD[ \sigma_{i-1}^z \sigma_{i}^z,\sigma_{i}^x \sigma_{i+1}^x] \nonumber \\
%	&\quad +Jh[ \sigma_{i-1}^z \sigma_{i}^z,(\sigma_i^x+\sigma_{i+1}^x)]+JD[\sigma_{i-1}^x \sigma_{i}^x,\sigma_{i}^z \sigma_{i+1}^z] \nonumber\\
%	&\quad +D^2[\sigma_{i-1}^x \sigma_{i}^x, \sigma_{i}^x \sigma_{i+1}^x]+Dh[\sigma_{i-1}^x \sigma_{i}^x,(\sigma_i^x+\sigma_{i+1}^x)] \nonumber\\
%	&\quad +Jh[(\sigma_{i-1}^x+\sigma_i^x),\sigma_{i}^z \sigma_{i+1}^z]+Dh[(\sigma_{i-1}^x+\sigma_i^x),\sigma_{i}^x \sigma_{i+1}^x] \nonumber \\
%	&\quad +h^2[(\sigma_{i-1}^x+\sigma_i^x),(\sigma_i^x+\sigma_{i+1}^x)]\big)\nonumber\\
%	&=-i\big(JD(\sigma_{i-1}^z[\sigma_i^z,\sigma_i^x]\sigma_{i+1}^x+\sigma_{i-1}^x[\sigma_i^x,\sigma_i^z]\sigma_{i+1}^z) \nonumber \\
%	& \quad +Jh(\sigma_{i-1}^z[\sigma_i^z,\sigma_i^x]+[\sigma_i^x,\sigma_i^z]\sigma_{i+1}^z)\big)\nonumber\\
%	&= JD(\sigma_{i-1}^z \sigma_i^y \sigma_{i+1}^x-\sigma_{i-1}^x \sigma_i^y \sigma_{i+1}^z)+Jh(\sigma_{i-1}^z \sigma_{i}^y - \sigma_i^y \sigma_{i+1}^z)
%\end{align}


\begin{thebibliography}{10}

\bibitem{Dutta}
A. Dutta, G. Aeppli, B. K. Chakrabarti, U. Divakaran, T. F. Rosenbaum and D. Sen, Quantum Phase Transitions in Transverse Field Spin Models: From Statistical Physics to Quantum Information (Cambridge University Press, Cambridge, 2015).
\bibitem{Frontiers}
A. M. Zagoskin, E. Ilichev, M. Grajcar, J. J. Betouras, and F. Nori, Front. in Physics {\bf 2}, 33 (2014).
\bibitem{Qureshi}
M. A. Qureshi, J. Zhong, J. J. Betouras, and A. M. Zagoskin, Phys. Rev. A 95, 032126 (2017).
\bibitem{Pfeuty}
 P. Pfeuty, Annals of Physics, {\bf 57}, 79 (1970).
\bibitem{Lieb}
 E. Lieb, T. Schultz and  D. Mattis, Annals of Physics, {\bf 16}, 407 (1961). 
\bibitem{Vidal1}
 G. Vidal, Phys. Rev. Lett., {\bf 98}, 070201 (2007).
\bibitem{White93}
S. R. White, Phys. Rev. B {\bf 48}, 10345 (1993).
\bibitem{Schollwoeck}
 U. Schollw\"{o}ck, Rev. Mod. Phys., {\bf 77}, 259 (2005).
\bibitem{McCulloch}
 I.P. McCulloch, arXiv:0804.2509 [cond-mat.str-el] (2008).
  \bibitem{Kjall}
  J. A. Kj\"all, M. P. Zaletel, R. S. K. Mong, J. H. Bardarson and F. Pollmann, Phys. Rev. B, {\bf 87}, 235106 (2013).
 \bibitem{DeChiara}
 G. De Chiara, L. Lepori, M. Lewenstein, and A. Sanpera, Phys. Rev. Lett. {\bf 109}, 237208 (2012).
 \bibitem{Latorre}
 J. I. Latorre, E. Rico and G. Vidal, Quantum Info. Comput., {\bf 4}, 48 (2004).
\bibitem{Antal}
 T. Antal, Z. R\'{a}cz, and L. Sasv\'{a}ri, Phys. Rev. Lett. {\bf 78}, 167 (1997).
 \bibitem{Eisler1}
 V. Eisler, Z. R\'{a}cz and F. van Wijland, Phys. Rev. E {\bf 67}, 056129 (2003).
 \bibitem{Kadar}
 Z. K\'{a}d\'{a}r and Z. Zimbor\'{a}s, Phys. Rev. A {\bf 82}, 032334 (2010).
\bibitem{Zotos}
 X. Zotos, F. Naef and P. Prelov\v{s}ek, Phys. Rev. B, {\bf 55}, 11029 (1997).
\bibitem{Torre}
E. G. Dalla Torre, E. Demler, T. Giamarchi and E. Altman,  Nature Physics {\bf 6}, 806 (2010).
\bibitem{Baruch}
E. Baruch and B. McCoy, Phys. Rev. A {\bf 3}, 786 (1971).
\bibitem{Calabrese}
 P. Calabrese and J. Cardy, J. Stat. Mech., P06002 (2004).
\bibitem{pollmann09}
F. Pollmann, S. Mukerjee, A. M. Turner and J. E. Moore, Phys. Rev. Lett., {\bf 102}, 255701 (2009).
\bibitem{tagliacozzo}
L. Tagliacozzo, T. R. de Oliveira, S. Iblisdir, and J. I. Latorre, Phys. Rev. B {\bf 78}, 024410 (2008).
\bibitem{Vidal2}
G. Vidal, J. I. Latorre, E. Rico and A. Kitaev, Phys. Rev. Lett., {\bf 90}, 227902 (2003).
\bibitem{Latorre2}
J. I. Latorre and A. Riera, J. Phys. A: Math. Theor. {\bf 42}, 504002 (2009).
\bibitem{Jafari}
 R. Jafari, Phys. Rev. B, {\bf 84}, 035112 (2011).
 \bibitem{Eisler2}
 V. Eisler and Z. Zimbor\'{a}s, Phys. Rev. A {\bf 71}, 042318 (2005).
 \bibitem{Keating}
 J. P. Keating and F. Mezzadri, Phys. Rev. Lett. {\bf 94}, 050501 (2005).
 \bibitem{Feldman}
K. E. Feldman, J. Phys. A: Math. Gen. {\bf 39}, 1039 (2006).
\bibitem{Genovese}
G. Genovese, Physica A {\bf 434}, 36 (2015).
\bibitem{Eggert}
 S. Eggert, I. Affleck, Phys. Rev. B, {\bf 46}, 10866, (1992).
\bibitem{Biella}
 A. Biella, A. De Luca, J. Viti, D. Rossini, L. Mazza, and R. Fazio, Phys. Rev. B {\bf 93}, 205121 (2016). 
\bibitem{Wei1}
T.-C. Wei, S. Vishveshwara, and P. M. Goldbart, Quantum Information and Computation {\bf 11}, 0326-354 (2011).
\bibitem{Wei2}
T.-C. Wei, D. Das, S. Mukhopadyay, S. Vishveshwara, and P. M. Goldbart, Phys. Rev. A {\bf 71}, 060305(R) (2005).
%\bibitem{Ma}
 %Y.Q. Ma and S. Chen, Phys. Rev. A {\bf 79}, 022116 (2009)

\end{thebibliography}
\end{document}